\documentclass{article}

\usepackage[final,nonatbib]{nips_2018}
\usepackage[utf8]{inputenc} 
\usepackage[T1]{fontenc}    
\usepackage{hyperref}       
\usepackage{url}            
\usepackage{booktabs}       
\usepackage{amsfonts}       
\usepackage{nicefrac}       
\usepackage{microtype}      
\usepackage{amsmath}
\usepackage{graphicx}
\usepackage{comment}

\usepackage[english]{babel}
\usepackage[utf8]{inputenc}
\usepackage{algorithm}
\usepackage[noend]{algpseudocode}

\title{FADL:Federated-Autonomous Deep Learning for  Distributed Electronic Health Record}
\author{
  Dianbo Liu ,Timothy Miller, Raheel Sayeed and Kenneth D. Mandl\thanks{Corresponding author: \texttt{Kenneth.Mandl@childrens.harvard.edu}} \\
  Computational Health Informatics Program\\
  Boston Children's Hospital and Harvard Medical School\
  ,One Autumn Street Boston, MA USA, 02215 \\
}

\begin{document}

\maketitle
\begin{abstract}
 Electronic health record (EHR) data is collected by individual institutions and often stored across locations in silos. Getting access to these data is difficult and slow due to security, privacy, regulatory, and operational issues. We show, using ICU data from 58 different hospitals, that machine learning models to predict patient mortality can be trained efficiently without moving health data out of their silos using a distributed machine learning strategy. We propose a new method, called Federated-Autonomous Deep Learning (FADL) that trains part of the model using all data sources in a distributed manner and other parts using data from specific data sources. We observed that FADL outperforms traditional federated learning strategy and conclude that balance between global and local training is an important factor to consider when design distributed machine learning methods , especially in healthcare. 
\end{abstract}

\section{Introduction}

Electronic Health Record (EHR) data, patient generated health data from mobile devices and other health related information are valuable for improving health outcomes, especially for precision medicine\cite{kohane2015ten,beam2016translating}. However, there are many challenges in utilizing these data efficiently. One of them is data access. Healthcare records are stored in different locations and data silos, including but not limited to hospitals, pharmacies, payors, and personal devices\cite{Goldstein2017OpportunitiesReview,mohammed2010centralized,bhatt2017internet,islam2015internet}. Traditionally, healthcare data distributed across sites centralized in a database for access for analysis \cite{liebowitz2017actionable,holzinger2016machine,hashem2015rise}. However, healthcare data transfers are complex because of strict regulations and sensitivity of the data \cite{mandel2016smart}. These hurdles not only make data utilization expensive but also slow down information flow in healthcare where timely updates are often important. 
\par
\begin{figure}
  \centering
   \includegraphics[scale=0.35]{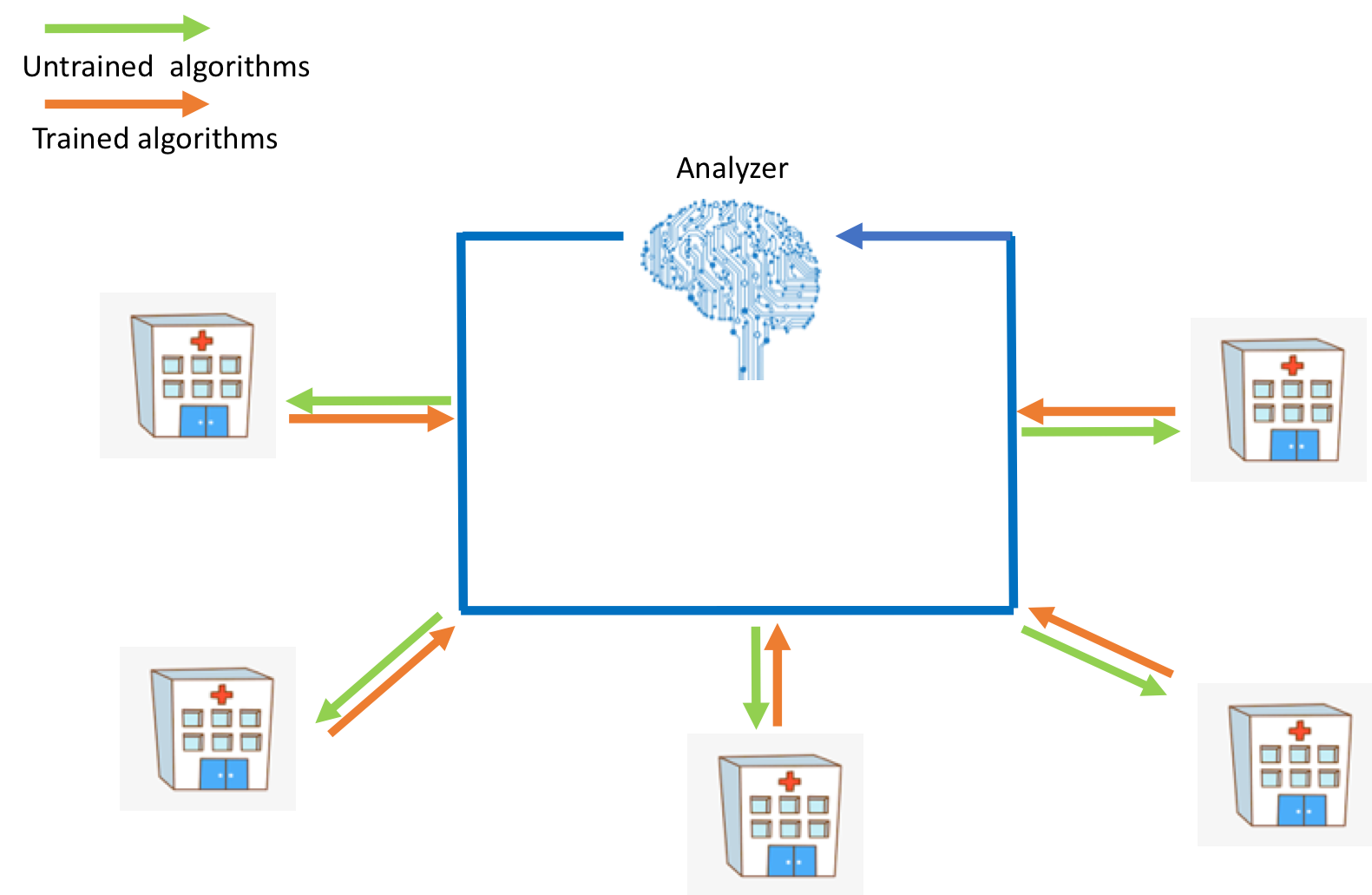}
\caption{Federated machine learning allows machine learning model be trained from multiple data sources without moving the data.}
\label{Federated}
\end{figure}

 
The process of using supervised machine learning for data analysis can be roughly divided into model training, where some datasets are used to optimize the model parameters, and prediction, where a trained model is used to make predictions on unseen data\cite{bishop2012pattern}. 
%
The motivation for federated or distributed machine learning is to train algorithms on different data sources in a distributed manner and aggregate the learned models (Figure \ref{Federated})\cite{konevcny2016federated,McMahan2016Communication-EfficientData,konecny2016federated}. 
%
In this paradigm, the algorithms that can learn from parts of the data are sent to each of the data sources for distributed training.
Parameters of all the locally trained models are then sent back to the analyzer to build a new ensembled model. This cycle repeats for a certain number of iterations. The machine learning model can be designed in such a way that it will not be possible to retrieve individual-level data of patients from the model.  Data-providing nodes retain health data within their institutional walls through this federated information flow.  We used hospital ICU data as an example to demonstrate how federated machine learning can train models using data unevenly distributed on multiple sources, and propose a new method called Federated-Autonomous learning that balances global and local training.

\section{Dataset, study cohort and methods }
The eICU Collaborative Research Database was developed by the by Philips eICU program\cite{pollard2018eicu} and populated with data from 208 critical care units throughout the continental United States for patients admitted in 2014 and 2015. We included data from 58 hospitals comprising 1,264,89 ICU admissions with discharge status information (alive or expired) (Table \ref{Cohort}). We developed a model that takes as inputs, the medications taken in the first 24 hours to predict mortality during the ICU admission. The cohort was administered 1400 different medicines in total. Binary information of whether a patient took each of the medicines in the first 24 hours after admission were used as input features. 5.5\% of patients died during the stay. We used 70\% of the ICU admissions for training, 10\% as validation set, and 20\% as test set. 

\begin{table}
  \caption{Study cohort (count by admission) }
  \label{Cohort}
  \centering
  \begin{tabular}{lll}
    
    \cmidrule(r){1-3}
    Information    & Alive     & Deceased \\
    \midrule
     Age&  &    \\
     0-18& 4269 & 403   \\
     18-60&48777  & 1945   \\
     >60 & 66867 &  4558  \\
    Gender&  &    \\
    Female& 54286 &  3094  \\
    Male&  65286& 3789   \\
    Medical information&  &    \\
    Length of stay in hrs, mean(std)&65.5(92.6)  & 98.3(162.5)   \\
    Number of drug started in the first 24 hours & 12.9(10.0) & 13.6(9.2)   \\

    \bottomrule
  \end{tabular}
\end{table}

\subsection{Neural network model for ICU mortality prediction }
In order to make a binary prediction of mortality of each ICU admission, we built a 3-layer fully connected artificial neural network model with 500, 100, and 1 neurons in corresponding layers using ReLU activation at hidden layers  and sigmoid activation at the output layer. Cross-entropy was used as the loss function for training. Patient mortality was used as the binary label with 0 for alive and 1 for deceased.  Each layer was L2 regularized with $\lambda$=0.01. The first training configuration, representing an upper bound, trains a centralized model in the traditional way, as if the ICU data from 58 hospitals can be moved to the same centralized database.

\begin{figure}
  \centering
  \includegraphics[width=0.7\linewidth]{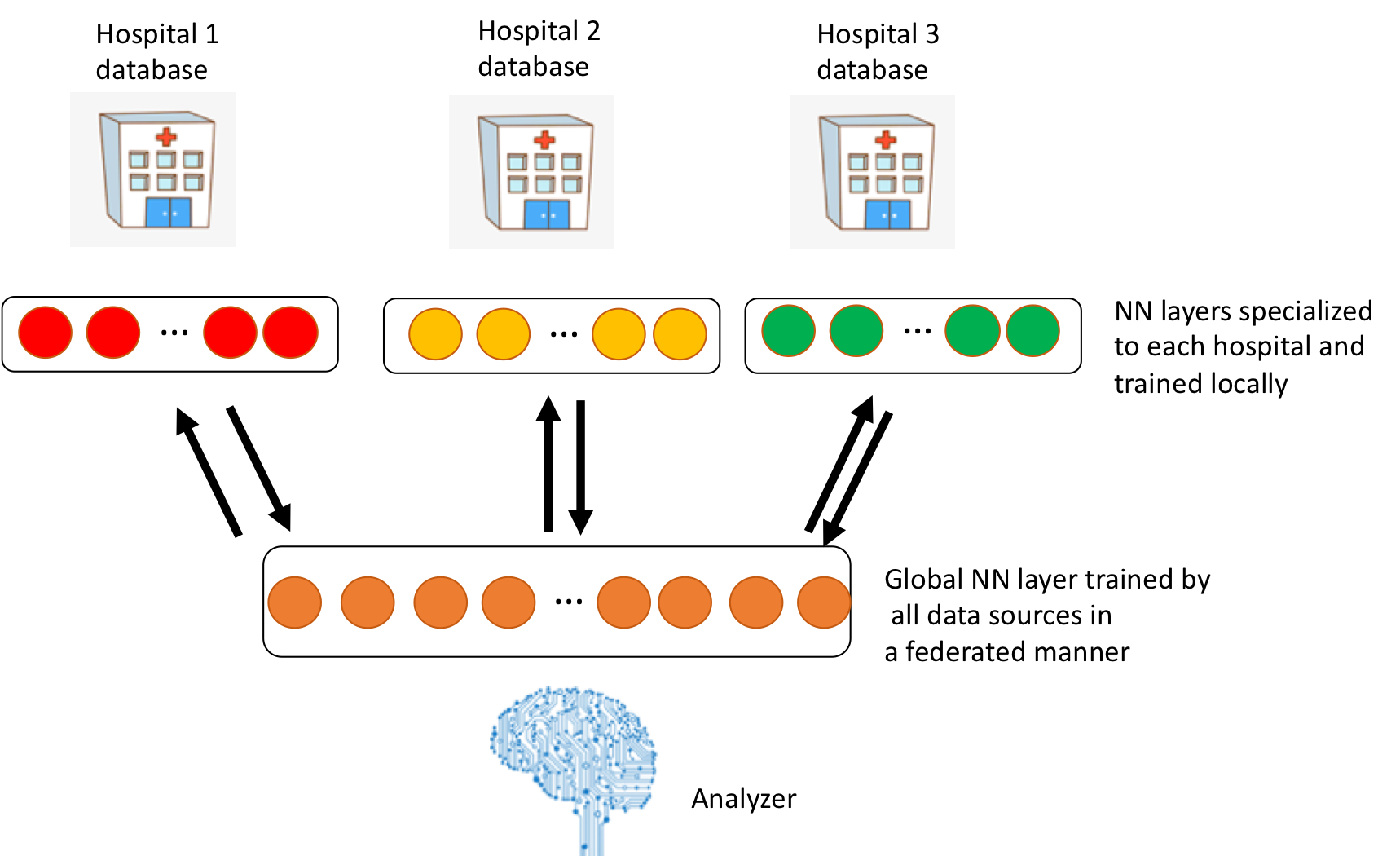}
  \caption{Federated autonomous deep learning (FADL) . Parts of the neural network were trained globally in a federated manner using all data sources. Other parts of the neural networks are specialized in each data source.}
  \label{FADL}
\end{figure}
\subsection{Federated learning on distributed hospital data}

Next, to mimic the real world medical setting, we assume all the ICU data from each of the 58 hospital stays at its local hospital. To train the deep learning model, we sent out models with identical parameters to all simulated hospitals nodes. The models were trained locally within each hospital using only data from that hospital. Parameters of the models were then sent back to the analyzer for aggregation, by averaging the parameters weighted by sample size \cite{konevcny2016federated,McMahan2016Communication-EfficientData,konecny2016federated}. After model aggregation, the updated model was sent out to all hospitals again to repeat the global training cycle.
Formally, the weight update is specified by:
\begin{equation}
W^{t}_{ag}=\sum_{i=1}^{k} \frac{n_i}{N} W^{t}_{i}
\end{equation}

where $W_{ag}$ is the combined parameter, $k$ is the number of data sources, $n_i$ is the number of admissions in the $i^{th}$ data source,  $N$ is the total number of admissions, and $W_i$ is the parameters learned locally from the $i^{th}$ data source. $t$ is the global cycle number  in the range of [1,T]. The objective function (cross-entropy) for this federated learning algorithm is:

\begin{equation}
\underset{f}{arg \space min} -\sum_{j=1}^{N}[y_j log f(x_j)+(1-y_j)log(1-f(x_j))])
\label{fig:obj}
\end{equation}

Where $x_{j}$ is the feature vector with dimension $(1,1400)$ and $y$ is the binary mortality label. $f$ is the neural network model. 
\subsection{Federated-Autonomous learning}
From the objective function in Equation~\ref{fig:obj}, It can be intuitively understood that the aim of the 
of the original federated learning algorithm is to minimize the average classification errors across all the data sources (hospitals) using information from all sources simultaneously. However, when there are a large number of data sources with different amounts of data of different properties, it may be difficult to balance what the model learns globally with locally-relevant information from each data source.
To tackle this challenge, we propose an FADL strategy where the first half of the neural network is trained globally using data from all sources and the second half of the neural network is trained locally to specialize in each data source (Figure \ref{FADL}). The algorithm is designed as in algorithm \ref{alg:FADL}. The same number of global cycles and local epochs were used for FADL and original federated Learning. It is worth emphasizing that both global and local training were conducted in a distributed manner , therefore, no data aggregation is needed.

\section{Results}
As the ICU mortality data are largely imbalanced, we used both AUCROC and AUCPR as measurements of accuracy\cite{buckland1994relationship,davis2006relationship}. When we conducted the model training in a centralized manner, the model was trained for 30 epochs with batch size=100. The model achieved AUCROC of 0.79 and AUCPR 0.21 on the test data (Table \ref{Performance}). 
\par 

When assuming data are stored in a distributed manner and can not be moved for centralized model training, we first trained the neural network using original federated learning. The same model architecture was used for federated learning as centralized learning. The model was trained for 20 global cycles involving all the 58 data sources. In each global cycle, the model was trained on each data source for 5 epochs before aggregation. The model trained using original federated learning achieved AUCROC of 0.75  and AUCPR of 0.16.  Next, we trained the model using a FADL strategy. FADL consist of two stages. In the first stage, all layers of the model were trained distributed across all sources, as in federated learning, for 10 global cycles with 5 local epochs each cycle. In the second stage, the parameters of the first neural network layer are fixed and only the 2nd and 3rd layers were trained for 50 epochs on each data source to generate 58 different models specialized for different hospitals. The first layer of the 58 models are identical and the 2nd and 3rd layers are different among models. When predicting mortality, each specialized model was used for corresponding hospital. Models trained using FADL perform at an AUCROC of 0.79 and AUCPR of 0.23 ,which is similar to our centralized learning model and superior to the federated learning model (Table \ref{Performance}).

\begin{table}
  \caption{Performance of centralized, original federated and Federated-autonomous learning }
  \label{Performance}
  \centering
  \begin{tabular}{lll}
    
    \cmidrule(r){1-3}
    Training method     & AUCROC     & AUCPR \\
    \midrule
    Centralized learning & 0.79 & 0.21    \\
    Original federated learning     & 0.75 & 0.16    \\
   Federated autonomous deep learning (FADL)     & 0.79      & 0.23  \\
    \bottomrule
  \end{tabular}
\end{table}

\begin{algorithm}
\caption{Federated autonomous learning algorithm}\label{alg:FADL}
\begin{algorithmic}[1]
\Procedure{FADL}{$X,Y,T_1,E_2$}\Comment{features,labels, stage1 and stage2 length}
\State Initialize weight $w_0$ of NN model $f(x)$
\State $t\gets 0$

\For {t in $1...T_1$}\Comment{First stage}
    \For{i in $1...K$} \Comment{Parallel training, K is the number of data sources}
        \State train $ f(x)$ and obtain $w^K_t$ 
    \EndFor
\State $w_t\gets \sum_{i=1}^{K} \frac{n_k}{N}w^K_t$

\EndFor
\State Freeze the first layer of the neural  \Comment{Second stage}
 \For{i in $1...K$} \Comment{Parallel training}
    \State $f^i(x) \gets f'(x)$ \Comment{$f'(x)$ is the trained model from first stage}
    \State Train 2nd and 3rd layers of $f^i(x)$ on data source K for $E_2$  epochs
 \EndFor

\State \textbf{return} $f^i(x),i \in [1...K]$
\EndProcedure
\end{algorithmic}
\end{algorithm}
\section{Conclusion}
We proposed a distributed neural network training method that balances global model training utilizing all data sources and local specialization that trains part of the model specifically on one data source. We showed that our FADL strategy outperformed traditional  federated learning and had similar accuracy to centralized learning. The balance between global and local learning is an important factor to consider when designing distributed machine learning methods,especially on health data\cite{liu2017deepfacelift,Rudovic2018PersonalizedTherapy}.

\bibliographystyle{thesnumb}
\bibliography{Mendeley,Extra}

\end{document}